\documentclass[useAMS,usenatbib,usegraphicx]{mn2e}
\usepackage{lscape}
\newcommand\KAB{K_{\rm AB}}

\newcommand\beff{b_{\rm eff}}

\newcommand\zb{z_{\rm B}}
\newcommand\Ks{K_{\rm s}}
\newcommand\Mmin{M_{{\rm min}}}
\newcommand\zp{z_{\rm p}}
\newcommand\Mh{\langle M_{h} \rangle}
\newcommand\Msun{{\rm M_{\odot}}}

\title[The very wide-field $gzK$ galaxy survey -- I]{The very wide-field $gzK$ galaxy survey -- I. Details of the clustering properties of star-forming galaxies at $z \sim 2$}
\author[Ishikawa et al.]{Shogo Ishikawa$^{1,2}$\thanks{E-mail: shogo.ishikawa@nao.ac.jp}, Nobunari Kashikawa$^{1,2}$,  Jun Toshikawa$^{2}$, and Masafusa Onoue$^{1,2}$\\
$^{1}$Depertment of Astronomy, School of Science, SOKENDAI (The Graduate University for Advanced Studies), Mitaka, Tokyo 181-8588, Japan\\
$^{2}$Optical and Infrared Astronomy Division, National Astronomical Observatory of Japan, Mitaka, Tokyo 181-8588, Japan}
\begin{document}

\pagerange{\pageref{firstpage}--\pageref{lastpage}} \pubyear{2015}

\maketitle

\label{firstpage}

\begin{abstract}
We present the results of clustering analysis on the $z \sim 2$ star-forming galaxies. 
By combining our data with data from publicly available archives, we collect $g$-, $\zb / z$-, and $K$-band imaging data over 5.2 deg$^{2}$, which represents the largest area BzK/gzK survey. 
We apply colour corrections to translate our filter-set to those used in the original BzK selection for the gzK selection. 
Because of the wide survey area, we obtain a sample of 41,112 star-forming gzK galaxies at $z \sim 2$ (sgzKs) down to $\KAB < 23.0$, and determine high-quality two-point angular correlation functions (ACFs). 
Our ACFs show an apparent excess from power-law behaviour at small angular scale $(\theta \la 0.01^{\circ})$, which corresponds the virial radius of a dark halo at $z \sim 2$ with a mass of $\sim 10^{13} \Msun$. 
We find that the correlation lengths are consistent with the previous estimates over all magnitude range; however, our results are evaluated with a smaller margin of error than that in previous studies. 
The large amount of data enables us to determine ACFs differentially depending on the luminosity of the subset of the data. 
The mean halo mass of faint sgzKs $(22.0 < K \leq 23.0)$ was found to be $\Mh = (1.32^{+0.09}_{-0.12}) \times 10^{12} h^{-1} \Msun$, whereas bright sgzKs ($18.0 \leq K \leq 21.0)$ were found to reside in dark haloes with a mass of $\Mh = (3.26^{+1.23}_{-1.02}) \times 10^{13} h^{-1} \Msun$. 
\end{abstract}


\section{INTRODUCTION}
Exploring galaxy formation and evolution is one of the most important issues in modern astronomy. 
The structure formation model suggesting that more massive objects in the Universe were rapidly assembled, and star formation activities were completed in an early epoch is known as ``downsizing'' \citep[e.g.,][]{cowie99}; however, it is not clear when and how the specific precesses in this evolutionary model occurred. 
The physical processes between baryons in the dark matter halo are so complex that it is not known how the luminous objects that we are able to observe directly were formed in the highly dense dark matter halo. 

To approach these issues, it is important to determine properties of galaxies, including the dark halo mass, the stellar mass, the star-formation rate (SFR), and the morphology over cosmic time, and to compare these quantities with theoretical models. 
The dark halo mass of galaxies is a particularly important parameter to trace the mass assembly history of galaxies because, according to the $\Lambda$CDM model, dark haloes grow monotonically with cosmic time by merging, irrespective of the baryon processes. 
However, measuring the mass of the dark halo is not straightforward. 
One of the most effective methods to determine the mass of dark haloes is to use the galaxy clustering strength. 
Various studies have revealed the dark halo mass in the distant Universe \citep[e.g.,][]{kashik06,hilde09,jose13} as well as in the local Universe \citep[e.g.,][]{zehavi05,zheng07}; however, this method requires a large number of samples in order to measure the clustering strength of galaxies, which has been a problem for $z \sim 2$ galaxies. 

The distinctive spectral features of galaxies at $z \sim 2$ are red-shifted from optical wavelengths to near-infrared (NIR) wavelengths. 
Wide-field NIR observation using ground-based telescopes is complicated because of the limited physical size of the NIR detectors and (with the exception of some atmospheric windows) poor sky transparency, as well as strong OH airglow emission lines and thermal emission. 
For this reason, wide-field galaxy survey data are lacking at $z \sim 2$ (this commonly referred to as the ``redshift desert''). 
Recent observations have shown that $z \sim 2$ is an important era in galaxy formation and evolution. 
For example, star formation in galaxies peaked at $1 < z < 2$ \citep[e.g.,][]{hopkins06,burgarrela13}, galaxy mass rapidly assembled at $1 < z < 3$ \citep[e.g.,][]{arnouts07,perez08,santini09,ilbert10}, and the number density of QSOs peaked at $z \sim 2$ \citep[e.g.,][]{palanque13}. 

A powerful method to select unbiased data with high completeness $z \sim 2$ galaxies has been proposed, namely the ``BzK selection'' technique \citep{daddi04}. 
The BzK selection method requires only $B$-, $z$-, and $K$-band photometric data, and enables us to select both star-forming BzKs (hereafter sBzKs) and passively evolving BzKs (hereafter pBzKs) simultaneously in a single colour--colour diagram. 
However, a precise clustering analysis of BzK galaxies requires a large number of galaxy samples and has not yet been performed due to the difficulties in obtaining wide-field $K$-band imaging data. 

A number of studies have attempted to reveal the clustering properties of BzK galaxies. 
\citet{kong06} constructed BzK galaxy samples over 1200 arcmin$^{2}$. 
They conducted the clustering analysis of sBzKs and pBzKs and concluded that pBzKs are more strongly clustered than sBzKs, suggesting that the morphology--density relation seen in the local Universe must already have been in place at $z \sim 2$. 
\citet{hayashi07} carried out a deep BzK galaxy survey in the Subaru Deep Field to investigate the properties of faint sBzKs and reported that the clustering strength of sBzKs depends on the $K$-band luminosity. 
\citet{mccracken10} succeeded in constructing a large number of BzK galaxy samples in the COSMOS field over 1.9 deg$^{2}$ down to $K<23.0$ and carried out a precise clustering analysis. 
Wide survey area and multiwavelength data from the COSMOS survey enabled them to determine the comoving correlation length for both sBzKs and pBzKs, respectively. 
More recently, \citet{bielby14} reported the clustering properties of $z \sim 2$ galaxies by splitting the galaxy samples into star-forming galaxies and passive galaxies based on the photometric redshift and galaxy colour. 
They revealed that passive galaxies are more strongly clustered than are star-forming galaxies; however, very massive star-forming galaxies ($M_{\star} \sim 10^{11} \Msun$) exhibited comparable clustering strength to passive galaxies. 
Their clustering analysis of star-forming galaxies at $z \sim 2$ was, however, confined to massive galaxies because of a shortage of galaxy samples at $z \sim 2$. 
\citet{bethermin14} succeeded in calculating the dark halo mass of sBzKs for wide mass range $(10^{10.4} \Msun < M_{\star} < 10^{11.4} \Msun)$ using both galaxy clustering and HOD analyses in the COSMOS field. 
The dark halo masses determined using both methods were in good agreement and were  consistent with the predictions of semi-analytic models. 
They calculated the dark halo mass at $z \sim 2$; however, the clustering analysis used had a poor signal-to-noise (S/N) ratio, especially at the large angular scale ($\theta \ga 0.1^{\circ}$), which made it difficult to determine the dark halo mass accurately. 

In this paper, we report a largest-ever wide-field galaxy survey over $\sim 5$ deg$^{2}$ down to $\KAB < 23.0$ using $g$-band data instead of $B$-band data (we term our sample galaxies ``gzKs'') and performed the clustering analysis of star-forming gzKs (hereafter sgzKs) with a significantly higher S/N ratio than has been reported previously. 
Because of the large number of sgzK samples, we are able to divide our sgzK samples into subsamples according to the $K$-band luminosity, and we investigate the clustering properties of each subsample. 
Our very wide survey area makes it possible to constrain the dark halo mass because we were able to determine an accurate angular correlation function at large angular scales, which is essential to determining the dark halo mass accurately. 
In this paper (paper I), we describe the measurement of the angular correlation function of star-forming galaxies at $z \sim 2$ based on this large sample, and the dark halo mass is determined using angular correlation functions. 
Additionally, we carried out a halo occupation distribution (HOD) analysis on the sgzKs, and we determine the distribution of galaxies in the dark haloes for each $K$-band limiting magnitude, which will be described in paper II. 

The framework of this paper is organized as follows. 
In Section 2, we give details of the data that were used as well as the method employed to construct the $K$-selected and sgzK catalogues. 
We describe the method used to correct for differences in the filter transmissions between our $g \zb K$ filters and the VLT-$BzK$ filters \citep{daddi04}. 
In Section 3, we describe the method to conduct clustering analysis using the angular correlation function. 
Taking advantage of the large number of sgzK samples, we resample the sgzK samples into subsamples according to the $K$-band magnitudes to investigate the clustering properties. 
The results of our clustering analysis of sgzKs are discussed in Section 4, and we give a conclusion in Section 5. 
All magnitudes and colours are in the AB system. 
Throughout this paper, we assume flat lambda cosmology $(\Omega_{{\rm m}} = 0.3$, $\Omega_{\Lambda} = 0.7)$, the Hubble constant is $h = {\rm H_{0}}/100 \, {\rm km \, s^{-1} Mpc ^{-1}} = 0.7$, and the normalization of the matter power spectrum is $\sigma_{8} = 0.8$. 
Assuming these cosmological parameters, the age of the Universe at $z = 2$ is $\sim 3.22$ Gyr, and 1 arcsec corresponds to 8.37$h^{-1}$ kpc in the comoving scale. 

\section{DATA}
\subsection{Photometric Data}
We collected wide-field imaging data by using both our own data and publically available archival data. 
Photometric data describing the $K$-band were obtained from the archives of the United Kingdom Infra-Red Telescope (UKIRT) Deep Sky Survey \citep[UKIDSS;][]{law07} in the Deep Extragalactic Survey (DXS). 
The UKIDSS data were acquired using the Wide Field Camera (WFCAM), which is composed of four $2,048 \times 2,048$ pixel detectors, with a separation of $12.'8$ between detectors. 
Each detector covers $13.'65 \times 13.'65$ of the sky at a pixel scale of $0.''187$ pixel$^{-1}$. The central wavelength and the full-width at half-maximum (FWHM) of the $K$-band of the WFCAM are $\lambda_{\rm {c}}=2.2 {\rm \mu m}$ and ${\rm FWHM} = 0.34 {\rm \mu m}$, respectively. 

We obtained $\zb$-band photometric data during the course of another observational program \citep{kashik15} using the Subaru Telescope/Suprime-Cam from June 22 to June 24, 2009. 
Suprime-Cam is a wide-field imaging instrument covering $34' \times 27'$ of the sky at a pixel scale of $0.''22$ pixel$^{-1}$. 
The $\zb$-filter ($\lambda_{\rm {c}}=8,842 {\rm \AA}, {\rm FWHM} = 689 {\rm \AA}$) is a custom-made filter that divides the SDSS $z$-band filter at $9,500 {\rm \AA}$ and their bluer one \citep{shima05}. 
One of our observation field ``VIMOS4'' is centered on $(22^{{\rm h}}20^{{\rm m}}00^{{\rm s}}, +00^{\circ}42'00''; {\rm J2000.0})$ and covers $\sim$3 deg$^{2}$ \citep{law07}. 
Here, 3 deg$^{2}$ of the VIMOS/DXS field was covered by $13$ Suprime-Cam field-of-views (FoVs). 
We adopted a common dithering circle pattern comprising a full cycle of dithering consisting of $5$ pointings. 
The total integration time was $1,800$ seconds per FoV. 
The sky condition was very good, with a seeing size of $0.''6$. 
Photometric calibration was carried out using the spectroscopic standard stars GD153 and Feige110. 
The data were reduced using the pipeline software package SDFRED \citep{yagi02,ouchi04a}. 
The package includes bias subtraction, flat fielding, a correction for image distortion due to the prime focus, PSF matching, sky subtraction, and mosaicking. 

We also retrieved $z$-band photometric data from the SMOKA data archive server \citep{baba02} to extend the survey field to 5.2 deg$^{2}$. 
These data satisfy the conditions that they connect continuously to our $\zb$-band field, and the limiting magnitude is comparable to our $\zb$-band data. 
The $z$-band data were obtained using the Subaru Telescope/Suprime-Cam. 

We retrieved $g$-band images, which cover the entire field where $K$ and $\zb/z$ images are available, using the Canada--France--Hawaii Telescope Legacy Survey \citep[CFHTLS;][]{gwyn11} archival data instead of using $B$-band. 
The CFHTLS data were acquired using the CFHT MegaCam, which covers $57.'6 \times 56.'4$ of the sky at a pixel scale of $0.''187$ pixel$^{-1}$, where the central wavelength and FWHM of the $g$-band of MegaCam are $\lambda_{\rm {c}}=4,870 {\rm \AA}$ and ${\rm FWHM} = 1,450 {\rm \AA}$, respectively. 

We collected $(g, \zb, K)$ bands of imaging data over 3.2 deg$^{2}$ and $(g, z, K)$ bands over 2.0 deg$^{2}$. 
The survey field of this study covers 5.2 deg$^{2}$, comprising the above two regions with different imaging datasets. 

\citet{fang12} showed that star-forming and passive galaxies at $z \sim 2$ are also able to be selected using the $(g-z)$ and the $(z-K)$ colours, which is similar to BzK colour selection. 
They constructed sgzK and pgzK samples by modifying the original BzK colour-selection criteria and validated their criteria using the stellar population synthesis model \citep{bc03}. 
We applied a similar gzK selection process to our data, although colour correction from our $(g-\zb)$ data to the VLT-$(B-z)$ data is required, as only the $\zb$-band is available. 

\begin{table}
\centering
\caption{Limiting magnitudes and covering areas of each band}
\begin{tabular}{ccc}
\hline\hline
 & limiting magnitude & area \\ 
 & (3$\sigma$, 2$'' \phi$ in AB) & (deg$^{2}$) \\ \hline
$g$ & 26.08 & 5.2 \\
$\zb$ & 25.55 & 3.2 \\
$z$ & 25.36 & 2.0 \\
$K$ & 23.31 & 5.2 \\
\hline \hline
\label{tab:limitmag}
\end{tabular}
\end{table}

\subsection{$K$-selected Catalogue}
The $g$- and $\zb/z$-band images were matched to the geometry of the $K$-band image using the IRAF task {\tt geomatch}. 
Object detection and photometry were carried out using SExtractor, version 2.8.6 \citep{bertin96}. 
Following object detection in the $K$-band, we used SExtractor in ``double-image mode'' to measure the object fluxes in the $g$- and $\zb$/$z$-bands with $2''$ aperture photometry. 
The size of the unit image was $\sim 4,170 \times 4,170$ pixels, which corresponds to $\sim 27'.8 \times 27'.8$. 
We carefully conducted masking on the low-S/N regions, such as near the edges of the images and close to the saturated objects, to remove objects with an uncertain flux. 
The number of masks was 690 over the field, and the effective survey area was 4.8 deg$^{2}$.  
We detected 339,581 objects in our field after masking. 

The limiting magnitudes were measured with a 3$\sigma$, 2$''$ aperture in AB magnitude. 
We used the IRAF task {\tt limitmag} in SDFRED to determine the limiting magnitudes and applied averaged data as limiting magnitudes in each band. 
Table \ref{tab:limitmag} lists a summary of the limiting magnitudes and the areas covered by each band. 
These limiting magnitudes were almost identical ($\delta$ mag $\sim 0.08$) over the entire field. 

\subsection{Band Corrections}
\begin{figure}
\centering
\includegraphics[width=80mm]{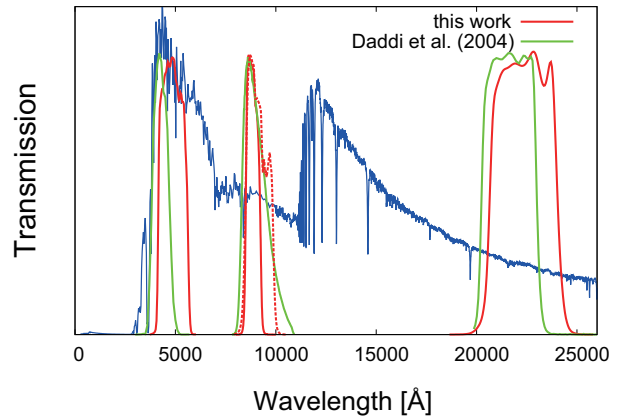}
\caption{Differences of the filter transmission profiles between our $g \zb K$-filters (red) and the VLT-$BzK$ filters (green) used in \citet{daddi04}. Our filters are composed of CFHT/MegaCam $g$-band, Subaru/Suprime-Cam $\zb / z$-band, and UKIRT/WFCAM $K$-band. Suprime-Cam $z$-band is presented by the dashed line. The blue line represents the typical SED of sgzK galaxies. }
\label{fig:band-daddi}
\end{figure}

We used the BzK selection method to select $z \sim 2$ galaxy samples; however, the filter set of our data differed slightly from the VLT-$BzK$ filter set \citep{daddi04}. 
Figure \ref{fig:band-daddi} shows a comparison of the transmission of our $g\zb K$-filters with that of VLT-$BzK$ filters. 
When adapting our $g \zb K$ data to the VLT-$BzK$ data, careful correction for differences in the filters is essential to adequately select $z \sim 2$ galaxies and to compare the results. 

\citet{daddi04} used the $\Ks$-band \citep{skrut06}, whereas we used $K$-band imaging data from the UKIDSS archive. 
We used the following conversion equation \citep{carp01} between these filters: 
\begin{equation}
\centering
\Ks = K + 0.002 + 0.026 \times (J - K).
\label{eq:Ks-conv}
\end{equation}
$J$-band imaging data were also retrieved from the UKIDSS archive, as both $K$-band photometric data and $J$-band magnitudes were available. 
To simplify the notation, we hereafter describe the $\Ks$-band magnitudes converted from $K$-band magnitudes using equation (\ref{eq:Ks-conv}) as ``$K$''. 
Following conversion of the $K$-band magnitudes, we limited our $K$-selected catalogue to $K=23.0$, which corresponds to a 70$\%$ completeness cut (see Section 3.3). 
We extracted all galaxies from our catalogue using the BzK diagram (see Section 2.4), and sorted them in magnitude bins from $K=16.0$ to $K=23.0$ with $\delta K = \pm 0.25$. 

\begin{figure}
\centering
\includegraphics[width=80mm]{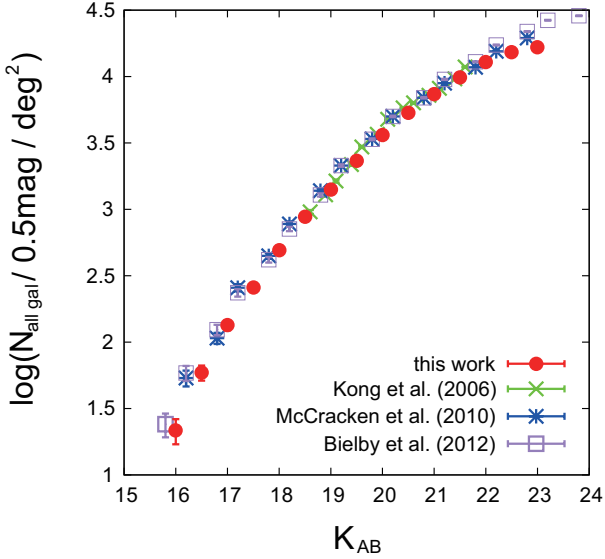}
\caption{Number count of the $K$-band selected galaxies after the correction of the detection completeness in our VIMOS4 field (red circles). Green, blue, and purple symbols represent the results of \citet{kong06}, \citet{mccracken10}, and \citet{bielby12}, respectively.}
\label{fig:Ncount_allgal}
\end{figure}

Figure \ref{fig:Ncount_allgal} shows a comparison of the number counts of our $K$-selected galaxies with previously reported studies. 
Our $K$-selected catalogue is consistent with previous works. 

The BzK selection technique selects $z \sim 2$ galaxies in the colour--colour diagram of $(B - z)$ versus $(z - K)$. 
We attempted to carry out colour conversion from $(g - \zb)$, $(g - z)$, and $(\zb - K)$ to $(B - z)$ and $(z - K)$ following $K$-band magnitude conversion. 
First, to correct for differences between the $\zb$-band and the $z$-band, we simultaneously measured the $\zb$- and the $z$-band fluxes for the stars in the $\sim 0.22$ deg$^{2}$ region, where photometric data are available for both, and derived the following conversion equation:  
\begin{equation}
\centering
(z - K) = 0.940 \times (\zb -K) + 0.00330
\label{eq:zb-conv}
\end{equation}
using a least-squares regression method. 

In the same way, we determined the following expressions for conversion: 
\begin{equation}
\centering
(g - z) = 0.900 \times (B - z)_{Subaru} - 0.120
\label{eq:gB-conv}
\end{equation}
and
\begin{equation}
\centering
(g - \zb) = 0.961 \times (g - z) + 0.115, 
\label{eq:zbz-conv}
\end{equation}
respectively. 
We employed the public photometric catalogue of the COSMOS survey, in which $g$-, $B$-, and $z$-band photometries are available, to derive the equation (\ref{eq:gB-conv}). 

\citet{mccracken10} introduced empirical relations to translate $(B -z)$ colour derived by the Subaru/Suprime-Cam data to $(B - z)$ colour derived by the VLT for objects with $(B - z)_{Subaru} < 2.5$, 
\begin{equation}
\centering
(B - z)_{VLT} = 1.0833 \times (B - z)_{Subaru} + 0.053
\label{eq:mc_lt2.5}
\end{equation}
and with $(B - z)_{Subaru} \geq 2.5$, 
\begin{equation}
\centering
(B - z)_{VLT} = (B - z)_{Subaru} + 0.27. 
\label{eq:mc_mt2.5}
\end{equation}
Using these empirical relations and equation (\ref{eq:gB-conv}), we obtain the relations that convert from the Subaru-$(g-z)$ colour to the VLT-$(B-z)$ colour for objects with $(g-z)<2.13$, 
\begin{equation}
\centering
(B-z)_{VLT} = 1.20 \times (g-z) + 0.198
\label{eq:conv_start}
\end{equation}
and with $(g-z) \geq 2.13$, 
\begin{equation}
\centering
(B-z)_{VLT} = 1.11 \times (g-z) + 0.404. 
\end{equation}

To convert from the Subaru-$(g-\zb)$ colour for objects with $(g-\zb)<2.16$, 
\begin{equation}
\centering
(B-z)_{VLT} = 1.25 \times (g-\zb) + 0.0535
\end{equation}
and with $(g-\zb) \geq 2.16$, 
\begin{equation}
\centering
(B-z)_{VLT} = 1.16 \times (g-\zb) + 0.271. 
\label{eq:conv_end}
\end{equation}

We validated these colour convention equations by using simulated galaxy spectral energy distribution (SED) models. 
The various SED models of galaxies were generated by varying physical properties such as the star-formation history, the SFR, the age, and the dust extinction using the stellar population synthesis models \citep{bc03}. 
We used the dust extinction law of \citet{calzetti00}. 
We determined the $(g-\zb)$ and $(\zb-K)$ colours from those SEDs, and translated them to the $(B-z)$ and $(z-K)$ colours using the relations presented above, and then traced the redshift evolution on the colour--colour diagram. 
We found that star-forming galaxies and passively evolving galaxies at $1.4 \la z \la 2.5$ with our band corrections met the original selection criteria proposed by \citet{daddi04}. 

\subsection{gzK Selection Method}
We constructed the $z \sim 2$ star-forming galaxy samples by applying the BzK selection method. 
We term our BzK selection method the ``gzK selection'' method, and we select star-forming galaxy samples as sgzKs because we use $g$-band photometric data instead of $B$-band data. 

The criterion used to select the sBzK galaxies proposed by \citet{daddi04} is given by
\begin{equation}
\centering
(z-K) \geq (B-z)-0.2, 
\label{eq:daddi_sBzK}
\end{equation}
and that for pBzK is given by
\begin{equation}
\centering
(z-K) < (B-z)-0.2 \cap (z-K)>2.5. 
\label{eq:daddi_pBzK}
\end{equation}
We applied these original criteria after converting $(g, \zb, z, \Ks)$ photometric data to $(B, z, K)$ photometric data using equations (\ref{eq:zb-conv}) and (\ref{eq:conv_start}) $\sim$ (\ref{eq:conv_end}). 

\begin{figure}
\centering
\includegraphics[width=80mm]{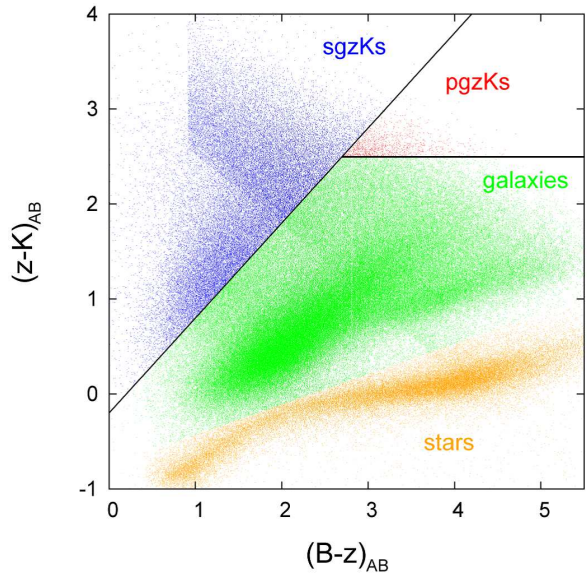}
\caption{A colour--colour diagram of our VIMOS4 field. The blue, red, green and orange dots represent sgzK galaxies, pgzK galaxies, other galaxies and stars, respectively. }
\label{fig:color-color}
\end{figure}

Figure \ref{fig:color-color} shows a colour--colour diagram for all the objects down to $K=23.0$ in our VIMOS4 field, where the blue and red dots represent the sgzKs and pgzKs, respectively. 
The number of selected sgzKs was N(sgzKs) = 41,112, and the number of pgzKs was N(pgzKs) = 1,313. 
Our $g$-band photometry was too shallow to detect all of the pgzK galaxies, which had very red $(g-z)$ colour (at least $g - z > 2.7$); therefore, we consider only the sgzKs galaxies. 
It should be noted that the $g$-band limiting magnitudes of non-detected objects in the $g$-band were replaced by the 2$\sigma$ limiting magnitude in the $g$-band, whereas non-detected objects in the $\zb/z$-band were removed from our catalogue because their locations in the colour--colour diagram were uncertain. 

\begin{figure}
\centering
\includegraphics[width=80mm]{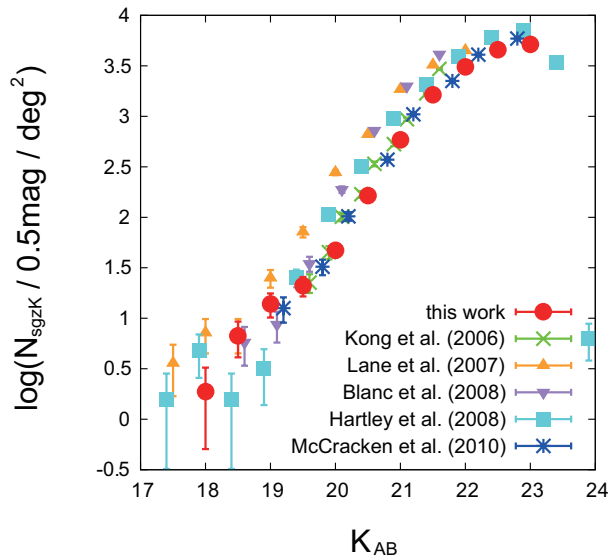}
\caption{Comparison of the number counts between our sgzK galaxies (red) after completeness correction and the sBzK galaxies of the previous studies. Green crosses, orange uptriangles, purple downtriangles, cyan squares and blue stars correspond to \citet{kong06}, \citet{lane07}, \citet{blanc08}, \citet{hartley08} and \citet{mccracken10}, respectively. }
\label{fig:Ncount_sgzK}
\end{figure}

Figure \ref{fig:Ncount_sgzK} shows a comparison of the number count between the sgzK and sBzK galaxies from our data and from previous studies. 
After correcting the completeness (see Section 3.2), our data were found to be consistent with almost all previous studies. 
Table \ref{tab:count_sgzK} lists a summary of the number count of our sgzKs. 
Our results were in agreement with almost all previous studies and were in particularly good agreement with the data reported by \citet{mccracken10}. 
Our study and \citet{mccracken10} are particularly in good agreement each other, while other previous studies have slightly higher number counts than our study. 
This might be caused by their small survey area. 

On the other hand, the bright end of our sgzK number count showed an excess compared with these data reported by \citet{blanc08} and \citet{hartley08}. 
Our very wide survey field enabled us to detect quite rare bright sgzKs at $K \la 19$. 

In our colour--colour diagram, there are some anomalous features. 
First, in the sgzK region, the sharp vertical line at $(B-z, z-K)$ $\sim$ $(1.0, 2.6)$ to $\sim (1.0, 3.3)$, which is originated from the non-detected objects in the $g$-band, can be seen. 
The z/$\zb$-band magnitudes of the objects on this line are so close to the limiting magnitude that the $(B-z)$ colour is nearly constant. 
We note that the $(B-z)$ colour was converted from original $(g-z)$ and $(g- \zb)$ colours by equation (7) and (9). 
In addition, a coherent line is seen from $(B-z, z-K)$ $\sim$ $(1.0, 2.6)$ to $\sim (4.0, 0.0)$ in the diagram. 
This line is caused by the limiting magnitude in $K$-band, making the constant $(B-K)$ colour. 

As shown in Figure \ref{fig:Ncount_sgzK}, the number count of our sgzK sample is consistent with the previous results, suggesting that these anomalous features do not largely affect the sgzK selection; however, $g$-band faint pgzKs, which show stronger correlation, especially at the small angular scale, than sgzKs \citep[e.g.,][]{mccracken10,zehavi11}, could come into the sgzK sample. 
To estimate the contamination of $g$-band faint pgzKs in our sgzK sample, we use the $K$-selected photometric redshift (photo-$z$) catalogue in the COSMOS field \citep{muzzin13}, which is approximately $1.0$ magnitude deeper than our sample. 
We extracted sgzKs and pgzKs from COSMOS photo-$z$ catalogue by applying our filter correction to that catalogue and the same colour selection criteria. 
Then we evaluated how many pgzKs are incorrectly selected as sgzKs by replacing $g$-band magnitudes by the 2$\sigma$ limiting magnitude when the objects were fainter than the 2$\sigma$ limiting magnitude. 
The contamination fractions of each limiting magnitude were found to be $f^{{\rm pgzK}}_{22<K<23} = 8.8\%$, $f^{{\rm pgzK}}_{21<K<22} = 8.8\%$, and $f^{{\rm pgzK}}_{18<K<21} = 6.0\%$, respectively. 
Therefore, we concluded that the contamination from pgzKs to our sgzK sample is not substantial. 

\begin{table}
\centering
\caption{The sgzK number count of each magnitude bin after completeness correction. }
\begin{tabular}{crc}
\hline\hline
$\KAB$ & N$_{{\rm sgzK}}$ & log(N$_{{\rm sgzK}}$) deg$^{-2}$ \\ \hline
18.0 & 9 & 0.273 \\
18.5 & 32 & 0.825 \\
19.0 & 67 & 1.143 \\
19.5 & 101 & 1.323 \\
20.0 & 225 & 1.673 \\
20.5 & 790 & 2.217 \\
21.0 & 2,801 & 2.766 \\
21.5 & 7,844 & 3.213 \\
22.0 & 14,789 & 3.489 \\
22.5 & 21,875 & 3.659 \\
23.0 & 24,704 & 3.712 \\ \hline \hline
\label{tab:count_sgzK}
\end{tabular}
\end{table}

\section{CLUSTERING ANALYSIS OF sgzK GALAXIES}
\subsection{Angular Correlation Function}
The angular correlation function (hereafter ACF) is a useful indicator of the clustering properties of galaxies. 
The ACF $\omega(\theta)$ is defined as the probability $P$ that a pair of galaxies exist within solid angles $\delta \Omega_{1}$, $\delta \Omega_{2}$ with a separation angle $\theta$ on the plane of the sky, and is given by
\begin{equation}
\centering
\delta P = \bar{n} [1+\omega(\theta)] \delta  \Omega_{1} \delta \Omega_{2}, 
\label{eq:prob_gal}
\end{equation}
where $\bar{n}$ represents the average surface density of galaxies.

We used the standard estimator proposed by \citet{landy93}, i.e., 
\begin{equation}
\centering
\omega(\theta) = \frac{DD -2DR + RR}{RR}, 
\label{eq:ls93}
\end{equation}
where $DD$, $DR$, and $RR$ are the number of galaxy--galaxy, galaxy--random, and random--random pairs within the angular range $\theta - \delta \theta /2 < \theta < \theta + \delta \theta /2$, respectively, and are normalized to the number of all galaxy--galaxy, galaxy--random, and random--random pairs. 
We note that $\theta$ is in units of degrees. 

We set the angular bin from log$(\theta) = -3.6$ to log$(\theta) = 0.2$, with a bin size $\delta$log$(\theta) = 0.2$ to determine the ACF for sgzKs galaxies. 
We generated 300,000 random points homogeneously over the survey field in order to reduce the Poisson error. 
The random points were distributed to avoid the masked region, which we used when obtaining sgzK galaxy samples. 

The ACF is generally given in power-law form, i.e., 
\begin{equation}
\centering
\omega(\theta) = A_{\omega} \theta^{1-\gamma}.
\label{eq:acf_power-law}
\end{equation}
We applied a fixed power-law gradient at large scales with $\gamma=1.8$ to compare the results with those of previous studies \citep[e.g.,][]{hayashi07,blanc08}. 

It is well known that the ACF reported by \citet{landy93} is biased due to the ``integral constraint'' caused by the limitations of the observation field. 
The integral constraint is given by 
\begin{equation}
\centering
C = \frac{1}{\Omega^{2}} \int_{\Omega} d\Omega_{1} d\Omega_{2} \omega(\theta).
\label{eq:IC_int}
\end{equation}

We can estimate the integral constraint as follows: 
\begin{equation}
\centering
C = \frac{\Sigma_{{\it i}} {\it RR_{i}} \omega(\theta_{i})}{\Sigma_{{\it i}} {\it RR_{i}}}, 
\label{IC_sum}
\end{equation}
provided that the ACF has already been obtained for each angular bin \citep{roche99}. 
The bias-corrected ACF is given by subtracting the integral constraint, i.e., 
\begin{equation}
\omega(\theta) = A_{\omega}(\theta^{1-\gamma} - C).
\label{eq:acf_ic}
\end{equation}
The integral constraint in the VIMOS4 field was $C=1.18$. 

The error in the ACF was estimated using the bootstrap method as follows. 
We randomly resampled our sgzK galaxies, allowing for redundancy and calculated the ACF at each step, repeating this 300 times. 
The uncertainty in each angular bin was determined from the root mean square of all of the bootstrap steps. 

\begin{figure}
\centering
\includegraphics[width=80mm]{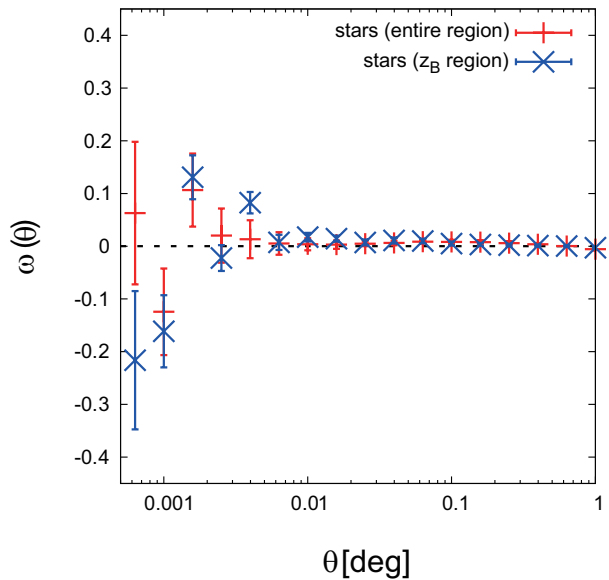}
\caption{The ACFs of stars (red lines, entire survey region; blue crosses, $\zb$-band covered region) in our survey field satisfying $18.0 \leq K \leq 23.0$. }
\label{fig:acf_star}
\end{figure}

To confirm the validity of the homogeneity, we check the ACF of stars in our sample. 
The distribution of stars can be regarded as random distribution; therefore, their ACF is expected to be almost . zero at all angular scale. 
Figure \ref{fig:acf_star} shows the ACF of stars. 
The amplitude of each angular bin is approximately zero, indicating that our sample are almost selected homogeneously. 

\subsection{Subsamples}
We divided our sample into subsamples according to the $K$-band luminosity to investigate the luminosity dependence of the clustering properties. 
Two subsample sets were created: a cumulative luminosity resampling subset and a differential luminosity resampling subset. 
The former was for use in comparing the results with previous studies, in which cumulative luminosity subsamples were created due to the small limited of sBzK data. 
We divided our large sgzK sample into five subsamples, i.e., $18.0 \leq K \leq 21.0$, $18.0 \leq K \leq 21.5$, $18.0 \leq K \leq 22.0$, $18.0 \leq K \leq 22.5$, and $18.0 \leq K \leq 23.0$. 
The sample number of each subsample are summarized in Table \ref{tab:data_cumu}. 

With the cumulative subsample set, however, it was difficult to describe the clustering properties of sgzKs by their $K$-band luminosity because cumulative resampling dilutes the differences among luminosity subsamples. 
Taking advantage of the large number of samples, we also divided our sample differentially according to the luminosity to assess the $K$-band luminosity dependence of sgzK clustering. 
We divided our sgzK samples into three subsamples based upon their $K$-band luminosity, i.e., $18.0 \leq K \leq 21.0$, $21.0 < K \leq 22.0$, and $22.0 < K \leq 23.0$, which allowed us to investigate how clustering and the physical properties of sgzK galaxies depend on the $K$-band luminosity.

\subsection{Redshift Distributions and Completeness}
We estimated the redshift distribution of our gzK-selected sample by applying the gzK selection method (including the filter corrections) to the $K$-selected COSMOS photo-$z$ catalogue \citep{muzzin13}. 
We assumed that the redshift distributions estimated from the COSMOS field ($\sim 1.9$ deg$^{2}$) were the same as our large $(\sim 5.2$ deg$^{2})$ field and that the cosmic variance was negligible. 

We used {\tt z\_m2}, which usually provides the most feasible $\zp$, as described in the EAZY manual \citep{bra08} in the COSMOS photo-$z$ catalogue. 
We extracted galaxies with a $K$-band magnitude in the range $18.0 \leq K \leq 23.0$ at $0<\zp<3$ and constructed a galaxy catalogue containing 79,284 galaxies. 
It should be noted that we excluded the galaxies with a large $\zp$ error, i.e., $(|{\rm {\tt zp} - {\tt z\_m2}}| > 0.5)$, where {\tt zp} is the best-fit redshift determined using SED fitting. 
The number of excluded galaxies was 318. 

We calculated the $(g-\zb)$ and $(\zb-K)$ colours using the best-fit SEDs and by estimating $\zp$ for each galaxy, and we derived the $g$- and $\zb$-band magnitudes according to the $K$-band magnitude from the COSMOS photo-$z$ catalogue. 
Artificial galaxies with assigned $(g, \zb, K)$ band photometries were then randomly distributed as point sources on the $g$-, $\zb$-, and $K$-band images. 
A total of 6,000 distributed galaxies were randomly selected from the galaxy catalogue allowing for redundancy in each limiting magnitude. 
Source detection and sgzK colour selection were carried out as with our data, and we obtained the redshift distributions for each luminosity subsample by repeating these processes 100 times and averaging the results of all steps.  

\begin{figure*}
\centering
\includegraphics[width=160mm]{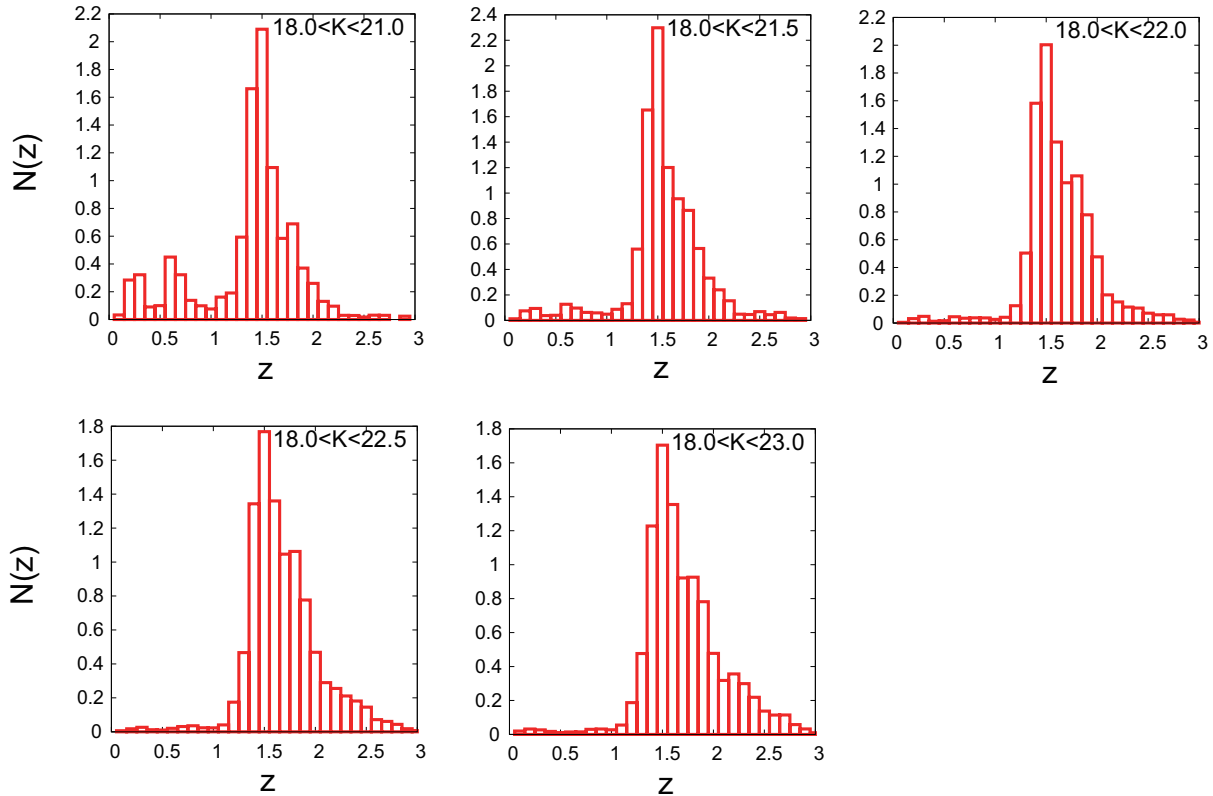}
\caption{Redshift distributions of each cumulative subsample. The mean number of redshift bin with $\delta z= 0.1$ width is normalized to one. We extracted the galaxies at $0 < z < 3$ from the COSMOS $K$-selected catalogue \citep{muzzin13} and applied gzK selection as we did on our own galaxy sample. }
\label{fig:Nz_cumu}
\end{figure*}

\begin{figure*}
\centering
\includegraphics[width=160mm]{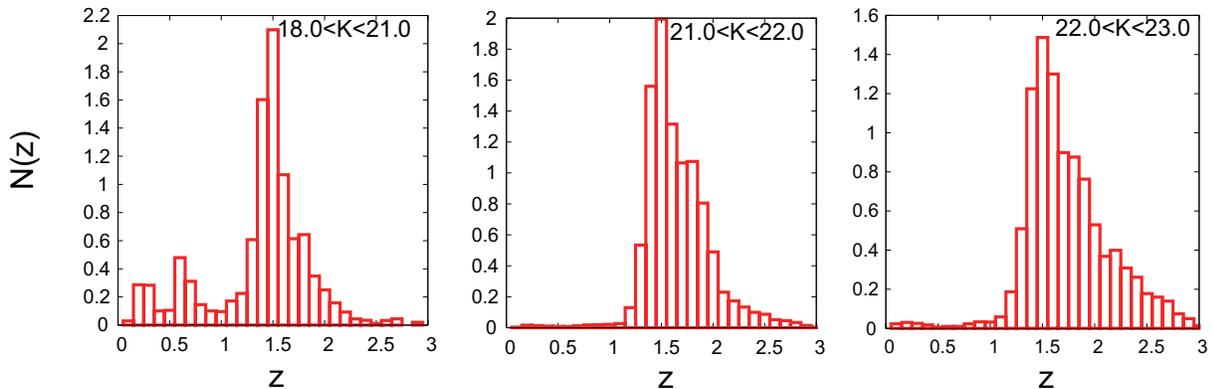}
\caption{Redshift distributions of each differential subsample. The method to derive these redshift distributions is the same as Figure \ref{fig:Nz_cumu}. }
\label{fig:Nz_diff}
\end{figure*}

Figure \ref{fig:Nz_cumu} and \ref{fig:Nz_diff} show the redshift distributions that satisfy the limiting magnitudes of our subsamples. 
The redshift range that satisfies the sgzK criterion is $1.4 \la z \la 2.5$, which indicates the validity of our colour corrections. 
Our redshift distributions are almost consistent with those of previous studies \citep[e.g.,][]{mccracken10,fang12}, though these are slightly shifted to lower $z$. 
This could be caused by the differences in the filters and the slight shallowness in the $K$-band of our sample. 
It should be noted that the brightest subsample ($18.0 \leq K \leq 21.0$) may be contaminated from low-$z$ galaxies. 
We corrected for the effects of this contamination in the ACF for this brightest subsample by multiplying $1/(1-f_c)^2$, where $f_{c}$ is the contamination fraction ($f_{c} \sim 0.2$), assuming that the contaminating sources are not clustered each other. 
It should be noted, however, that this assumption is strictly incorrect for contaminated low-$z$ galaxies, whose clustering amplitudes are difficult to be estimated. 
Other subsamples exhibited less contamination with low/high-$z$ galaxies ($f_c < 0.1$), and we did not apply a correction factor to these data. 

\begin{figure}
\centering
\includegraphics[width=80mm]{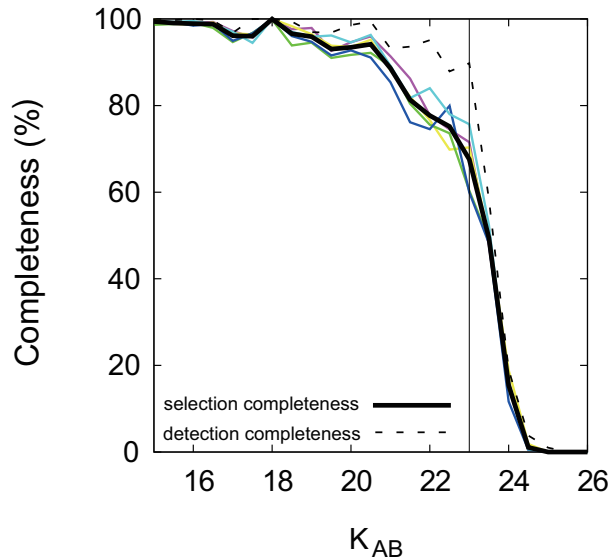}
\caption{Completeness of our sample. The coloured thin lines represent the results in different regions and the black solid thick line is the average of the results in 5 regions. The dotted line is the detection completeness, which represents how inputted sgzKs were detected as objects. We applied $18.0 \leq K \leq 23.0$ magnitude cut to our sample, corresponding to approximately 70$\%$ completeness cut. }
\label{fig:completeness}
\end{figure}

The redshift distribution that we estimated using the above procedure shows the completeness of our sample. 
Figure \ref{fig:completeness} shows our sample completeness; we obtained completeness of 70$\%$ with a limiting magnitude of $\KAB = 23.0$. 

\section{RESULTS AND DISCUSSIONS}
We compare the results of our clustering analysis based on the cumulative luminosity subsampling with those of previous studies. 
We then show the results of the differential luminosity subsamples to reveal the luminosity dependence of the clustering properties. 

\subsection{Cumulative Luminosity Subsample}
\subsubsection{Angular Correlation Functions of Cumulatively Resampled sgzK Galaxies}

\begin{figure}
\centering
\includegraphics[width=80mm]{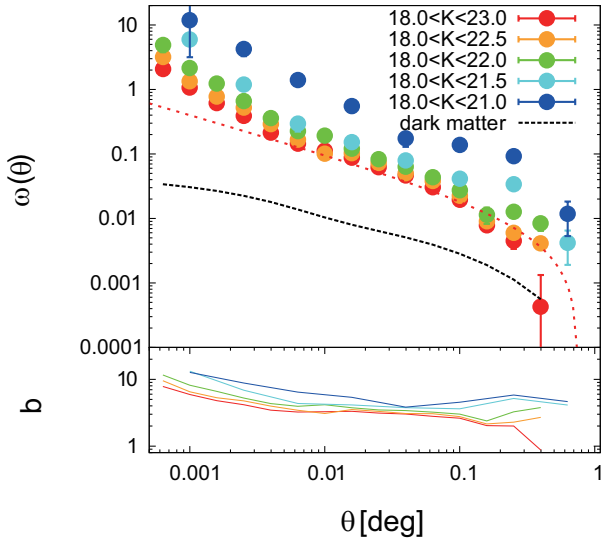}
\caption{Upper panel: ACFs of the cumulatively resampled sgzK galaxies. The limiting magnitudes of each sgzK subsample are $18.0 \leq K \leq 23.0$, $18.0 \leq K \leq 22.5$, $18.0 \leq K \leq 22.0$, $18.0 \leq K \leq 21.5$, and $18.0 \leq K \leq 21.0$ (red, orange, green, cyan, and blue circles). We confirmed the fact that more bright sgzK galaxies show the strongly clustering, reported by \citet{hayashi07}. The red-dotted line represents the result of single power-law fit to the ACF of the total sample at large angular scales to show the excess from a power law at small angular scales. The dotted line represents the model prediction of the dark matter ACF computed using the nonlinear power spectrum. Lower panel: The bias parameters of sgzKs. Bias parameter is defined as $b(\theta)=\sqrt{\omega_{{\rm sgzK}}(\theta) / \omega_{{\rm DM}}(\theta)}$.}
\label{fig:acf_cumu}
\end{figure}

The upper panel of Figure \ref{fig:acf_cumu} shows the ACFs of each limiting-magnitude subsample. 
Bright sgzK galaxies were more strongly clustered than were faint sgzKs, which is consistent with previous studies \citep[e.g.,][]{kong06,hayashi07,mccracken10}. 
The bin sizes of bright sgzK subsamples $(18.0 \leq K \leq 21.0$, $18.0 \leq K \leq 21.5)$ were doubled $(\delta$log$(\theta) = 0.4)$ to gain the S/N ratio. 
Because of the accurate ACFs due to the large number of sgzK samples, our ACFs show an apparent excess from the power law at small angular scale (i.e., $\theta \la 0.01^{\circ}$). 
These characteristics are attributed to the so-called ``1-halo term'', which comes from galaxy clustering in the same dark halo. 
$\theta \sim 0.01^{\circ}$ corresponds to $0.21\, h^{-1}$Mpc at $z \sim 2$ at a physical scale and is comparable to the virial radius $(\sim r_{200})$ of a dark halo at $z = 2$ with a mass of $\sim 10^{13} \Msun$, where the virial radius is $r_{200} \sim 0.12\, h^{-1}$Mpc. 
The virial radius of the dark halo was estimated using 
\begin{equation}
\centering
r_{200} = (\frac{G M_{200}}{100 H^{2}(z)})^{1/3}, 
\label{eq:virial}
\end{equation}
where $G$ is the gravitational constant, $M_{200}$ is the virial mass of the dark halo \citep{carroll92,ferguson04}, and $H(z)$ is the Hubble parameter with a redshift $z$ as given by 
\begin{equation}
\centering
H(z) = H_{0}[\Omega_{m}(1+z)^{3} + (1-\Omega_{m}-\Omega_{\Lambda})(1+z)^{2} + \Omega_{\Lambda}]^{1/2}. 
\label{eq:hubble}
\end{equation}
On the other hand, the ACFs at large angular scale (i.e., $\theta \ga 0.01^{\circ}$) were well approximated by a power law with an index of $\gamma = 1.8$, referred to as the ``2-halo term'', which originates from galaxy clustering in different dark haloes. 
These two components can be accurately described by the halo occupation distribution (HOD) model, which predicts the galaxy distribution in the dark halo as a function of the halo mass. 
We will describe the results of HOD analysis of our sgzK galaxies in our paper II. 

We also show the model prediction of the dark matter ACF at $z \sim 2$ using the selection function of sgzKs with a limiting magnitude of $18.0 \leq K \leq 23.0$. 
This dark matter ACF was computed using the nonlinear $\Lambda$CDM power spectrum \citep[][and the references therein]{hamana04}. 

The lower panel of Figure \ref{fig:acf_cumu} shows the bias parameters of the sgzKs, which are defined as
\begin{equation}
\centering
b(\theta)=\sqrt{\frac{\omega_{{\rm sgzK}}(\theta)}{\omega_{{\rm DM}}(\theta)}}, 
\label{eq:bias}
\end{equation}
where $\omega_{{\rm sgzK}}(\theta)$ and $\omega_{{\rm DM}}(\theta)$ are the amplitudes of the ACFs of sgzKs and dark matter at the angular scale $\theta$, respectively. 
The values of our bias parameter were in the range $3 < b < 5$ at the large scale, which is consistent with the results of \citet{blanc08}. 
The bias parameters were larger at small angular scales than at large angular scales, indicating the excess from the power law. 

By applying the least $\chi^{2}$ fitting of the single power law (equation \ref{eq:acf_power-law}) with fixed $\gamma = 1.8$, we obtained best-fit values of the amplitude of the ACF at 1$^{\circ}$. 
We fitted the power law to the data for large angular scale $(\theta \ga 0.01^{\circ})$ where the 1-halo term is negligible. 
Table \ref{tab:data_cumu} lists the resulting amplitudes of the ACF; we can see that the clustering amplitude is dependent on the $K$-band luminosity. 

Our measurement of the amplitudes at $18.0 \leq K \leq 23.0$ was $A_{\omega}(1^{\circ})= (3.33 \pm 0.09) \times 10^{-3}$, which is larger than the value of $(1.79 \pm 0.17) \times 10^{-3}$ reported by \citet{hartley08}, as well as $(1.27 \pm 0.23) \times 10^{-3}$ reported by \citet{mccracken10}. 
For $18.0 \leq K \leq 22.0$, we found $A_{\omega}(1^{\circ})= (3.83 \pm 0.14) \times 10^{-3}$, which is larger than the value of $(3.14 \pm 1.12) \times 10^{-3}$ reported by \citet{blanc08} and $(2.12 \pm 0.65) \times 10^{-3}$ by \citet{mccracken10}. 
This deviation may be due to the fact that the survey area of the previous studies was too small to provide a high-quality signal at the large scale $(\theta > 0.1^{\circ})$. 
The ACFs reported in most previous studies were truncated at $\theta < 0.1^{\circ}$ or declined due to the effects of integral constraints. 
With our results, however, which are based on wide-field data, we calculated the ACF over a wide angular scale of $0.01^{\circ} < \theta < 0.5^{\circ}$, which enabled us to more accurately determine the amplitude, especially for the 2-halo term of the ACF. 
The amplitude of the ACF can be calculated from the large-scale galaxy clustering; the angular range was approximately $0.01^{\circ} \la \theta \la 0.1^{\circ}$, which makes it difficult to calculate the amplitude accurately if intermediate to large-scale clustering is not well determined. 

In addition, \citet{sato14} pointed out that the clustering amplitude reported by  \citet{mccracken10} was weaker than those of the other studies. 
\citet{sato14} also presented the correlation functions, which is inconsistent with the result of  \citet{mccracken10}, at the COSMOS filed, though the origin of this discrepancy was unclear. 

The results of our brightest three bins were in good agreement with those of \citet{kong06} although the error bars for their data were relatively large. 
\citet{blanc08} attributed the large amplitude reported by \citet{kong06} to the effects of cosmic variance; however, our results, where the survey field was more than 10 times larger than that of \citet{kong06}, are less affected by cosmic variance. 
For this reason, we can calculate the amplitude of the ACF of $K$-bright sgzKs more accurately than \citet{kong06} were able to. 

\subsubsection{Clustering in Real-Space}
The correlation amplitude of the ACF, $A_{\omega}$, can be transformed into the three-dimensional correlation length by assuming a redshift distribution, where the correlation length corresponds to the three-dimensional clustering strength. 
The redshift distributions we used are described in Section 3.2 (see Figure \ref{fig:Nz_cumu} and \ref{fig:Nz_diff}). 

The spatial two-point correlation function, $\xi(r)$, is described in power-law form as follows: 
\begin{equation}
\centering
\xi(r) = (\frac{r}{r_{0}})^{-\gamma}, 
\label{eq:2cf}
\end{equation}
where normalization factor $r_{0}$ is termed the correlation length, and $\gamma$ is the gradient of the power law. 
We assumed $\gamma=1.8$, in a similar way as with the ACF. 

The conversion between the ACF and the two-point correlation function was implemented using Limber's equation \citep{limber53}. 
Provided that the redshift distribution is known, the amplitude of the ACF can be translated to the correlation length as follows \citep{peebles80,efsta91}: 
\begin{eqnarray}
\centering
A_{\omega}({\rm 1rad}) &=& r_{0}^{\gamma} \sqrt{\pi} \frac{\Gamma(\gamma - 1/2)}{\Gamma(\gamma / 2)} \nonumber \\
 &\times & \frac{\int_0^{\infty} F(z) D_{A}^{1-\gamma}(z) N(z)^{2} g(z) dz}{[\int_{0}^{\infty} N(z) dz]^{2}}. 
\label{eq:limber}
\end{eqnarray}
The term $A_{\omega}({\rm 1 rad})$ corresponds to the amplitude of the ACF at $\theta = 1$ radian, where $\Gamma$ is the Gamma function, $D_{A}$ is the angular diameter distance, and $N(z)$ is the redshift distribution, respectively. 
The function $g(z)$ depends on the cosmological parameters and is given as follows: 
\begin{equation}
\centering
g(z) = \frac{H_{0}}{c} (1+z)^{2} \sqrt{1+\Omega_{m}z+\Omega_{\Lambda}[(1+z)^{-2}-1]} .
\label{eq:limber_g}
\end{equation}
$F(z)$ is a function describing the redshift evolution of the two-point correlation function $\xi(r)$, i.e., 
\begin{equation}
F(z) = (1+z) / (1+z_{c}) ^{-(3+\epsilon )}, 
\label{eq:limber_f}
\end{equation}
where $z_{c}$ is the center of the redshift distribution. 
We assumed $\epsilon$ as $\epsilon=-1.2$ \citep[see also][]{ouchi04b}. 

\begin{figure}
\centering
\includegraphics[width=80mm]{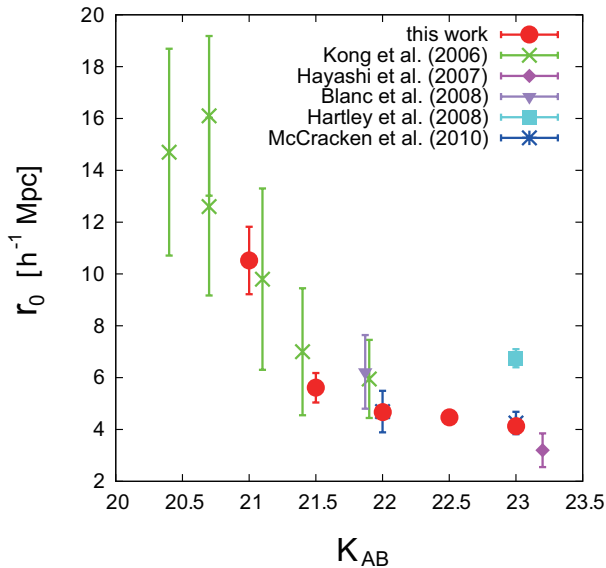}
\caption{Comparison of our correlation lengths of sgzKs with previous studies. Our results (red filled circles) are consistent with previous studies over all magnitude range and measured with small error bars. All correlation lengths are in units of $h^{-1}$Mpc, where $h=0.7$. }
\label{fig:r0_cumu}
\end{figure}

The correlation lengths are listed in Table \ref{tab:data_cumu}, and Figure \ref{fig:r0_cumu} shows a comparison of our results with those of previous studies. 
We found that brighter sgzKs have larger correlation lengths, which indicates that brighter galaxies reside in more massive haloes and exhibit stronger clustering. 
It should be noted that because of the large sample size, our results are characterized by smaller error bars than those of previous studies. 
Our results also show excellent agreement with previous studies over all magnitudes, with the exception of \citet{hartley08}. 
This discrepancy between the result of \citet{hartley08} and rest of the studies we compare may be caused by the inaccuracy of their sample selection \citep[see also][]{mccracken10,sato14}. 
As discussed in section~4.1.1, our ACFs have higher amplitudes than the previous studies, whereas our correlation lengths exhibit good agreement with the previous results. 
This may be due to the slight difference of the redshift distribution, that is, our sample has a redshift distribution shifted to lower-$z$ than the previous studies. 
We confirmed that the correlation length becomes smaller when the redshift distribution shifts low-$z$. 

The correlation length for sgzKs with $18.0 \leq K \leq 21.0$ were determined using the contamination-corrected ACF. 
Our correlation length of the brightest subsample is relatively large compared with the result of \citet{kong06}, which was not corrected for contamination; however, as shown in Figure \ref{fig:Nz_cumu}, the brightest subsample was more contaminated by the low/high-$z$ galaxies and should be corrected accordingly. 
The difference in the correlation lengths can be explained by the lack of correction for contamination in the data reported by \citet{kong06}. 

\subsection{Differential  Luminosity Subsample}
\begin{figure}
\centering
\includegraphics[width=80mm]{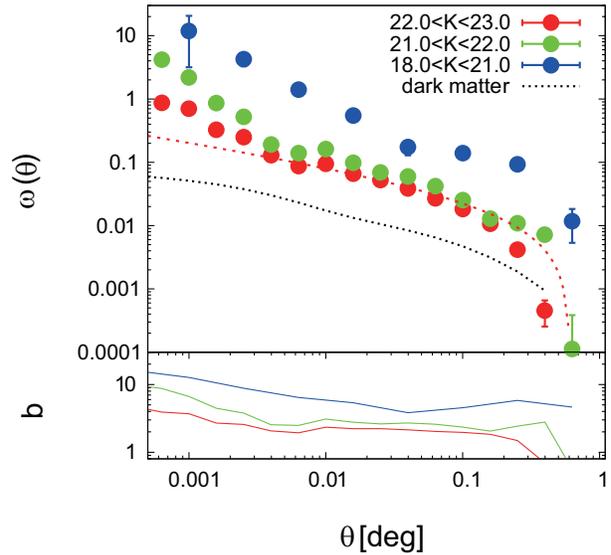}
\caption{Upper panel: ACFs of the differentially resampled sgzK galaxies. The limiting magnitudes of each sgzK subsample are $22.0 < K \leq 23.0$, $21.0 < K \leq 22.0$, and $18.0 \leq K \leq 21.0$ (red, green and blue circles). Lower panel: The bias parameters of sgzKs. }
\label{fig:acf_diff}
\end{figure}

In this section, we show the results of clustering analysis on subsamples with different luminosities. 
Figure \ref{fig:acf_diff} shows the ACFs of sgzKs with different luminosities. 
The bin size of the brightest sgzK subsample $(18.0 \leq K \leq 21.0)$ was increased to $(\delta$log$(\theta) = 0.4)$ to increase the S/N ratio. 
The error of each data point was larger than that for the cumulative subsamples due to the smaller sample size; however, these ACFs clearly show dependence on the $K$-band magnitudes and have the apparent excesses at small angular scale, as was the case for the cumulative subsamples. 

In the same manner as with the cumulative resampling, we fitted our sgzK subsamples using a power law to determine the amplitudes of the ACFs and the correlation lengths. 
These data are listed in Table \ref{tab:data_cumu}. 

\begin{figure}
\centering
\includegraphics[width=80mm]{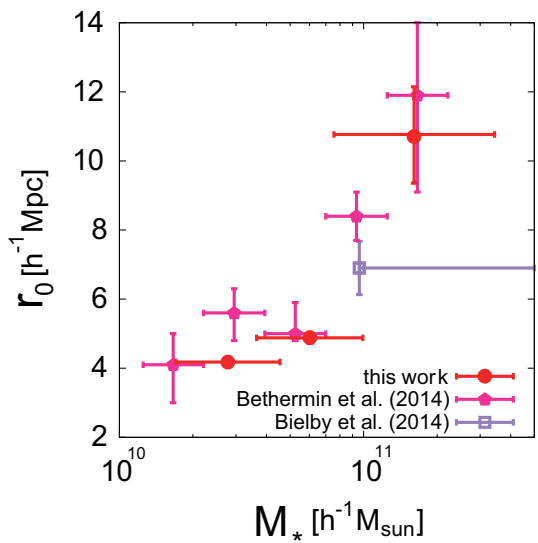}
\caption{The dependence of the correlation length upon the stellar mass of sgzKs/sBzKs. Our measurements (red) show the good agreement with \citet{bethermin14} and \citet{bielby14} (magenta and purple, respectively).  All correlation lengths are in units of $h^{-1}$Mpc, where $h=0.7$. }
\label{fig:r0-Mstar}
\end{figure}

\citet{bielby14} showed that the correlation length of a star-forming galaxy depends on the stellar mass, whereas the correlation length of a passive galaxy does not (or is only weakly dependent of the stellar mass), in combination with the results of \citet{coil08}, \citet{bielby10}, and \citet{mccracken10}. 
Figure \ref{fig:r0-Mstar} shows the relationship between the correlation length and the stellar mass for both this study and previous studies \citep{bielby14,bethermin14}. 
We note that the stellar masses of our samples are estimated by their $(z-K)$ colours and $K$-band magnitudes (see Section~4.3), whereas \citet{bethermin14} and \citet{bielby14} estimated by the SED fitting. 
The dependence of our correlation lengths on the stellar mass is consistent with previous studies. 

\subsection{Dark Halo Mass Estimation by the Large-Scale Clustering of sgzK Galaxies}
We calculated the dark halo mass residing in our sgzK samples using the results of our accurate clustering analysis. 
We used the analytical model proposed by \citet{sheth01b} and \citet{mowhite02}, which has been shown to provide a connection between the bias of the dark halo and the mass of the dark halo based upon ellipsoidal collapse \citep{sheth01a}, provided bias of dark halo and the redshift are given. 

We determined the amplitude of the two-point correlation function at $r=8h^{-1}$Mpc , $\xi_{{\rm sgzK}}(8 \, h^{-1}{\rm {\rm Mpc}})$, assuming a power-law form of equation (\ref{eq:2cf}) and the correlation lengths. 
We define the effective bias parameter $\beff$ as
\begin{equation}
\centering
\beff = \sqrt{\frac{\xi_{{\rm sgzK}}(8h^{-1} {\rm Mpc})}{\xi_{{\rm DM}}(8h^{-1}{\rm Mpc})}}, 
\label{eq:beff}
\end{equation}
where $\xi_{{\rm DM}}(8h^{-1}{\rm Mpc})$ is the model predicted correlation function of dark matter at an effective redshift with a scale of $8h^{-1}{\rm Mpc}$ of each subsample \citep[][and the references therein]{hamana02}. 

The bias of the dark halo can be calculated by the analytic formulae given by \citet{sheth01a} and \citet{mowhite02}. 
Here, we give a summary of these models. 
We define the linear growth factor $D(z)$ as
\begin{equation}
\centering
D(z) \equiv \frac{g(z)}{g(0) \, (1+z)}, 
\label{eq:lineargrowth}
\end{equation}
where $g(z)$ depends on the cosmological parameters, i.e., 
\begin{eqnarray}
g(z) & \approx & \frac{5}{2} \Omega_{{\rm m}}(z) [\Omega_{{\rm m}}^{4/7}(z) - \Omega_{\Lambda}(z) \nonumber \\
 & + & (1 + \Omega_{{\rm m}}(z)/2) (1 + \Omega_{\Lambda}(z)/70)]^{-1}. 
\label{eq:sheth_g}
\end{eqnarray}
$\Omega_{{\rm m}}(z)$ and $\Omega_{\Lambda}(z)$ are the cosmological parameters at a given redshift, and they satisfy
\begin{eqnarray}
\centering
\Omega_{{\rm m}}(z) &=& \frac{\Omega_{{\rm m}} \, (1+z)^{3}}{E^{2}(z)} \nonumber \\
\Omega_{\Lambda}(z) &=& \frac{\Omega_{\Lambda}}{E^{2}(z)}, 
\end{eqnarray}
where $E(z)$ is given by
\begin{equation}
\centering
E(z) = [\Omega_{\Lambda} + (1 - \Omega_{{\rm m}} - \Omega_{\Lambda}) (1 + z)^{2} + \Omega_{{\rm m}} (1 + z)^{3} ]^{1/2}. 
\label{eq:sheth_e}
\end{equation}
The bias of the dark halo with a mass of $M$, $b(M)$, is given by 
\begin{equation}
\centering
b(M) = 1 + \frac{1}{\delta_{c}} [\nu^{2} + b \nu^{2(1-c)} - \frac{\nu^{2}/\sqrt{a}}{\nu^{2c} + b(1-c)(1-c/2)}], 
\label{eq:bias_halo}
\end{equation}
where $\nu$ is defined by $\nu \equiv \sqrt{a} \delta_{c} / D(z) \sigma(M)$ and we assume the constants to be $a=0.707$, $b=0.5$, $c=0.6$, and $\delta_{c}=1.69$ \citep{sheth01a}. 
The term $\sigma(M)$ represents the root mean square of the density field fluctuation with a mass scale $M$, i.e., 
\begin{equation}
\centering
\sigma^{2} (M) = \frac{1}{2 \pi^{2}} \int^{\infty}_{0} k^{2} P(k) \tilde{W} (kR)^{2} dk
\label{eq:sheth_sigma}
\end{equation}
and the redshift dependence is included via the linear growth factor, $D(z)$. 
The term $P(k)$ is the matter power spectrum and has a power-law index of $n=1$. 
We calculated the power spectrum using the {\tt CAMB} package \citep{lewis00,challinor11}. 
The term $\tilde{W}$ is a window function, i.e., 
\begin{equation}
\centering
\tilde{W}(kR) = 3 \frac{{\rm sin}(kR) -kR{\rm cos}(kR)}{(kR)^{3}},
\label{eq:sheth_window}
\end{equation}
and $R$ is a Lagrangian radius of the halo, which gives the halo mass dependence, i.e., 
\begin{equation}
\centering
R(M) \equiv (\frac{3M}{4\pi \bar{\rho}_{0}})^{1/3},
\label{eq:sheth_radius}
\end{equation}
where $\bar{\rho}_{0}$ represents the current mean density of the Universe. 

We estimate the mean dark halo mass $\Mh$ and the minimum dark halo mass $M_{{\rm min}}$ from a comparison between the effective bias parameter and the bias of dark halo \citep[see also][]{hayashi07}. 
We define the mean dark halo mass when the bias of dark halo with a mass of $\Mh$ is equal to the effective bias, i.e., 
\begin{equation}
\centering
\beff = b(\Mh),
\label{eq:mean_mass}
\end{equation}
whereas the minimum dark halo mass is defined that the effective bias from equation (\ref{eq:beff}) is equal to another expression of the effective bias as, 
\begin{equation}
\centering
\beff = \frac{\int_{\Mmin}^{\infty} b(M) n(M) dM}{\int_{\Mmin}^{\infty} n(M) dM}, 
\label{eq:beff_hayashi}
\end{equation}
where $n(M) dM$ is the halo mass function. 

Table \ref{tab:data_cumu} lists the dark halo masses. 
Our method to calculate the dark halo mass is based on the wide survey area and can thus be expected to be accurate, as the results depend strongly on the clustering signals at large angular scales. 
Our measurements satisfy $\Mh \approx  3\Mmin$ over almost all limiting magnitudes, which is consistent with the results of \citet{hayashi07}, who reported that the mean dark halo mass is mainly determined by the less massive haloes, which are more numerous than the more massive ones. 
It should be noted that this measurement method of using large-scale clustering assumes that each dark halo must contain a single galaxy. 
This assumption may be erroneous, however, as massive haloes have been reported to contain multiple galaxies (e.g., galaxy groups/clusters in the local Universe), whereas less massive dark haloes may contain no galaxy at all. 
A more precise halo model, such as HOD, which describes the number of galaxies in the dark halo as a function of the mass of halo, may be beneficial to provide a more detailed description of the structure model. 

\citet{bielby14} and \citet{bethermin14} determined the dark halo mass by this method, though they divided their sample by the stellar mass. 
The $(z-K)$ colour and $K$-band magnitude allowed us to estimate the galaxy stellar mass by using the galaxy model of \citet[][and the references therein]{koyama13}. 
The conversion equation from the $(z-K)$ and $K$-band magnitudes of sgzKs to stellar mass is given by
\begin{eqnarray}
{\rm log(M_{\star} / 10^{11} M_{\odot})}& = & -0.4 \times (K - 21.90) \nonumber \\
& + &  (0.086 - 1.28 \times {\rm exp}(-0.921 \times (z-K))).  \nonumber \\
 & & 
\end{eqnarray}
We note that this was derived by assuming the Salpeter IMF \citep{salpeter55} and that the formation redshift is $z_{\rm f} = 5$. 
The scatter in the stellar mass of each galaxy was $\approx 0.3$ dex. 
We estimated the stellar mass of each sgzK using this equation and determined the average stellar mass, together with the standard deviation of each subsample. 

We find that the minimum mass of the dark halo that resides in sgzKs satisfying $18.0 \leq K \leq 23.0$ is $\Mmin = (4.60 \pm 0.38) \times 10^{11} h^{-1} {\rm M_{\odot}}$ and the mean halo mass is $\Mh = (1.23 \pm 0.10) \times 10^{12} h^{-1} \Msun$, which are approximately three times more massive than \citet{hayashi07} reported for $K < 23.2$. 
Additionally, \citet{blanc08} reported a minimum dark halo mass down to $K \la 22.0$ of $\Mmin \approx 3 \times 10^{12} h^{-1} \Msun$, which is also more massive than our estimation, i.e., $\Mmin = (8.45^{+1.75}_{-1.50}) \times 10^{11} h^{-1} \Msun$. 
These inconsistencies may be caused by the shallowness of their $K$-band photometry data and the small sample size. 
We also compared our results with those of \citet{bielby14} and \citet{bethermin14} by calculating the stellar mass of each subsample. 
Our results were in good agreement in terms of the mean halo mass reported by \citet{bielby14} and \citet{bethermin14} over the entire stellar mass range. 
However, \citet{bielby14} only estimated the dark halo mass of the massive galaxies, ${\rm M}_{\star} \sim 10^{11} h^{-1} \Msun$, and the error bars of the dark halo mass given by \citet{bethermin14} were relatively large. 
Here, however, the dark halo masses of $z \sim 2$ galaxies were determined for a wide range of $K$-band luminosities (and stellar masses), and we are able to provide relatively small error bars. 

As discussed above, we succeeded in calculating accurate dark halo masses using large-scale clustering via our high-quality ACFs; however, it is essential to discuss the relationship between galaxy stellar masses and dark halo masses to reveal the star-formation history of galaxies. 
In our forthcoming paper, we will discuss the evolution of the dark halo mass, the number of satellite galaxies in the dark halo, and the relationship between the dark halo mass and the stellar mass using our accurate dark halo mass data (including the HOD-derived dark halo mass and the galaxy stellar masses) to describe the evolution history of the galaxies. 

\section{CONCLUSIONS}
We have described the colour-correction method, the number count, and the clustering properties of sgzK galaxies. 
Using the $g$-band data instead of $B$-band data, we succeeded in applying the gzK selection method to a largest-ever wide field. 
Our survey area was 5.2 deg$^{2}$ (4.8 deg$^{2}$ after masking), which is $\sim 2.5$ times larger than the widest BzK galaxy survey reported previously \citep{mccracken10}. 

The conclusions of this paper can be summarized as follows. 
\begin{enumerate}
\item Based on wide-field observation, we collected 41,112 sgzK galaxies, which is approximately twice the number of sBzK galaxies reported by \citet{mccracken10}. 
Because of the large number of galaxy samples, we obtained the high-quality ACFs and report a more accurate clustering analysis compared with previous BzK clustering studies. 
Our ACFs show an apparent excess from the power law, which is indicated by the halo model. 
We resampled our large number of sgzK sample into cumulative luminosity subsamples and differential luminosity subsamples to investigate the dependence of sgzK clustering properties on the $K$-band magnitude. 

\item We found that the amplitudes of the ACFs determined using our cumulative luminosity subsamples were larger than those of previous studies. 
The amplitude of the ACFs is determined using the large-scale clustering of galaxies, and our ACFs have particularly large signals because of the wide survey area. 
This is the principal reason that the amplitudes of our ACFs may be considered more accurate that those of previous studies. 

\item We calculated correlation lengths using the amplitudes of the ACFs, which are in good agreement with the results of previous studies, except for \citet{hartley08}. 
The error bars of our correlation lengths are small compared with those of  previous studies because of the high S/N ratio. 

\item We found that correlation lengths calculated using our ACFs are dependent on the $K$-magnitude, i.e., brighter sgzKs in the $K$-band have a large correlation length, which is consistent with the results of many previous studies. 
It follows that bright sgzKs reside in massive dark haloes, which shows a agreement with the correlation between the correlation length and the dark halo mass inferred by \citet{hayashi07} and \citet{bethermin14}. 

\item The dark halo masses were estimated from the large-scale sgzK clustering with small error bars and confirmed that our dark halo masses are in agreement with those of  \citet{bielby14} and \citet{bethermin14}. 
\end{enumerate}

\section*{Acknowledgments}
We are obligated to Masao Hayashi to his useful advice on this study. 
We also thank to Yusei Koyama to give us the conversion equation of stellar mass by using his model. 
This work is based on data collected at the Subaru Telescope, which is operated by the National Astronomical Observatory of Japan (NAOJ). 
This research was supported by the Japan Society for the Promotion of Science through Grant-in-Aid for Scientific Research 23340050. 

This research is based in part on data obtained as part of the United Kingdom Infra-Red Telescope (UKIRT) Deep Sky Survey. 
This study is based on observations obtained with MegaPrime/MegaCam, a joint project of CFHT and CEA/DAPNIA, at the Canada--France--Hawaii Telescope (CFHT), which is operated by the National Research Council (NRC) of Canada, the Institut National des Sciences de l'Univers of the Centre National de la Recherche Scientifique (CNRS) of France, and the University of Hawaii. 
This work is based in part on data products produced at TERAPIX and the Canadian Astronomy Data Center as part of the Canada--France--Hawaii Telescope Legacy Survey, a collaborative project of NRC and CNRS. 


\begin{landscape}
\begin{table}
\centering
\caption{The detailed data of cumulative/differential subsamples}
\begin{tabular}{ccccccccc}
\hline\hline
 & N(sgzK) & $z_{{\rm eff}}$ & A$_{\omega}$(1$^{\circ}$) $\times$ 10$^{-3}$ & $r_{0} $ [$h^{-1}$ Mpc] & b$_{{\rm eff}}$ & $\Mmin \, [h^{-1} \Msun]$ & $\Mh \, [h^{-1} \Msun$] & $M_{\star} \, [h^{-1} \Msun$] \\ \hline
18.0$\leq K \leq$21.0 & 2,086 & 1.43 & 18.48 $\pm$ 4.57 & 10.52 $\pm$ 1.30 & 4.16 $\pm$ 0.46 & $(1.71^{+0.76}_{-0.60}) \times 10^{13}$ & $(3.26^{+1.23}_{-1.02}) \times 10^{13}$ & $(1.61^{+1.83}_{-0.85}) \times 10^{11}$ \\ 
18.0$\leq K \leq$21.5 & 6,243 & 1.61 & 7.23 $\pm$ 1.40 & 5.61 $\pm$ 0.57 & 2.36 $\pm$ 0.21 & $(1.84^{+0.97}_{-0.63}) \times 10^{12}$ & $(4.39^{+1.82}_{-1.45}) \times 10^{12}$ & $(1.24^{+1.00}_{-0.55}) \times 10^{11}$\\ 
18.0$\leq K \leq$22.0 & 15,247 & 1.68 & 4.36 $\pm$ 0.27 & 4.67 $\pm$ 0.17 & 2.00 $\pm$ 0.07 & $(8.45^{+1.75}_{-1.50}) \times 10^{11}$ & $(2.14^{+0.27}_{-0.31}) \times 10^{12}$ & $(8.64^{+7.00}_{-3.87}) \times 10^{10}$ \\
18.0$\leq K \leq$22.5 & 29,158 & 1.73 & 3.83 $\pm$ 0.14 & 4.47 $\pm$ 0.08 & 1.92 $\pm$ 0.03 & $(6.78^{+0.54}_{-0.50}) \times 10^{11}$ & $(1.78^{+0.14}_{-0.13}) \times 10^{12}$ & $(6.13^{+5.56}_{-2.92}) \times 10^{10}$\\
18.0$\leq K \leq$23.0 & 41,112 & 1.76 & 3.33 $\pm$ 0.09 & 4.12 $\pm$ 0.07 & 1.79 $\pm$ 0.03 & $(4.60^{+0.38}_{-0.38}) \times 10^{11}$ & $(1.23^{+0.10}_{-0.10}) \times 10^{12}$ & $(4.95^{+4.80}_{-2.44}) \times 10^{10}$ \\ \hline
18.0$\leq K \leq$21.0 & 2,086 &  1.43 & 18.48 $\pm$ 4.57 & 10.52 $\pm$ 1.30 & 4.16 $\pm$ 0.46 & $(1.71^{+0.76}_{-0.60}) \times 10^{13}$ & $(3.26^{+1.23}_{-1.02}) \times 10^{13}$ & $(1.61^{+1.83}_{-0.85}) \times 10^{11}$  \\ 
21.0$< K \leq$22.0 & 13,163 & 1.65 & 4.81 $\pm$ 0.25 & 4.88 $\pm$ 0.26 & 2.08 $\pm$ 0.06& $(1.07^{+0.15}_{-0.17}) \times 10^{12}$ & $(2.55^{+0.34}_{-0.32}) \times 10^{12}$ & $(6.02^{+3.91}_{-2.37}) \times 10^{10}$ \\ 
22.0$< K \leq$23.0 & 25,865 & 1.73 & 3.32 $\pm$ 0.09 & 4.18 $\pm$ 0.07 & 1.81 $\pm$ 0.03 & $(4.93^{+0.44}_{-0.43}) \times 10^{11}$ & $(1.32^{+0.09}_{-0.12}) \times 10^{12}$ & $(2.80^{+1.77}_{-1.08}) \times 10^{10}$ \\
\hline \hline
\label{tab:data_cumu}
\end{tabular}
\end{table} 
\end{landscape}

\end{document}